\documentclass[aps,prb,showpacs,twocolumn,amsmath,amssymb,superscriptaddress,letterpaper]{revtex4}
\usepackage{times}
\usepackage{amsfonts}
\usepackage{mathrsfs}
\usepackage{graphicx}% Include figure files
\usepackage{dcolumn}% Align table columns on decimal point
\usepackage{bm}% bold math
\usepackage{color}

\usepackage[colorlinks,bookmarks=false,citecolor=blue,linkcolor=red,urlcolor=blue]{hyperref}
\def\be{\begin{equation}}       \def\ee{\end{equation}}
\def\bea{\begin{eqnarray}}      \def\eea{\end{eqnarray}}

\begin{document}
\title{A possible new family of unconventional high temperature superconductors}

\author{Jiangping Hu}\email{jphu@iphy.ac.cn }
 \affiliation{Beijing National Laboratory for Condensed Matter Physics, and Institute of Physics, Chinese Academy of Sciences, Beijing 100190, China}
\affiliation{Collaborative Innovation Center of Quantum Matter, Beijing, China}
\affiliation{School of Physics, University of Chinese Academy of Sciences,  Beijing 100049, China}

\author{Congcong Le}
\affiliation{Beijing National Laboratory for Condensed Matter Physics, and Institute of Physics,
Chinese Academy of Sciences, Beijing 100190, China}

\begin{abstract}
We suggest a new family of Co/Ni-based materials that may host unconventional high temperature superconductivity (high-T$_c$). These materials carry layered square lattices with each layer being formed by vertex-shared transition metal tetrahedra cation-anion complexes. The electronic physics in these materials  is determined by the two dimensional layer and is fully attributed to the three near degenerated $t_{2g}$ d-orbitals close to a $d^7$ filling configuration in the d-shell of Co/Ni atoms . The electronic structure meets the necessary criteria for unconventional high T$_c$ materials proposed recently by us to unify the two known high-T$_c$ families, cuprates and iron-based superconductors. We predict that they host superconducting states with a d-wave pairing symmetry with T$_c$ potentially higher than  those of iron-based superconductors. These materials, if realized, can be  a fertile new ground to study strongly correlated electronic physics  and provide decisive evidence for superconducting pairing mechanism.
\end{abstract}


\maketitle

%In order to drive mechanism for high temperature superconductivity and technological application ,the search of new materials with high critical temperature is in high demand. In the past few decades,iron pnictides and cuprates with high critical temperature always generated considerable research interest\cite{Kamihara2008,Ren2008,Wang2008,Bednorz1986}.
%
%
%Here we present possible new superconductivity materials by first-principles calculations.
%Cuprates\cite{Bednorz1986}, first discovered in 1986, and iron-based %superconductors\cite{Kamihara2008,Ren2008,Wang2008},discovered in early 2008, are two classes of %unconventional high temperature superconductors.ever since understanding the superconducting %mechanism behind unconventional high temperature superconductors has become a great challenge in %condensed matter physics.Therefore,the search of new materials with high critical temperature is %in high demand.

% half iron, full t2g orbitals
%orbital, magnetic superconductivity
%briging cuprates, iron-based superconductors

Successful theoretical predictions of high temperature superconducting materials rarely happen. The two known families of high T$_c$ materials, cuprates\cite{Bednorz1986} and iron-based superconductors\cite{Kamihara2008-jacs}, were discovered accidentally without any theoretical guide. Theoretical studies have been mainly devoted to explain rich phenomena observed in experiments.    After almost three decades of intensive research, it has become extremely clear that if there is any chance to solve  the elusive high T$_c$ mechanism, a successful theoretical prediction of new high T$_c$ materials is necessary.

 Recently, we suggest that   a special electronic trait that separates the two high T$_c$ families  from other correlated electronic materials is that  in both high T$_c$ families, those d-orbitals that make the strongest in-plane d-p couplings in the cation-anion complexes   are isolated near Fermi surface energy\cite{Hu-tbp,Hu-genes, Hu-s-wave}. In magnetically-driven superconducting mechanism,  this property makes the effective antiferromagnetic(AFM) superexchange interactions to  maximize their contribution to superconducting pairing and simultaneously  reduces other unwanted side effects from other orbitals. We also further argued that  this property can only be realized in very limited special cases\cite{Hu-genes}. Realizing such a property requires a strict symmetry match between local building blocks and global lattices, as well as a specific electron filling configuration in the d-shells of transition metal atoms.
 In cuprates, which possess  perovskite or perovskite-like structures,  the speciality is only realized near the $d^9$ filling configuration in an  octahedra (or square) complex to isolate the e$_g$ $d_{x^2-y^2}$ orbital near Fermi energy. In iron-based superconductors, it is only realized near the $d^6$ filling configuration of a  tetrahedra complex to isolate two t$_{2g}$ d$_{xy}$ -type orbitals\cite{Hu-tbp,Hu-genes}. Therefore, this speciality allows us to explain why high T$_c$ is such a rare phenomenon.  It can be considered as a gene type character to guide us to search for or predict possible new high T$_c$ materials\cite{Hu-tbp}.

 Following the above analysis, we have predicted that the gene exists in a two dimensional hexagonal lattice formed by edge-shared trigonal biprymidal complexes with a $d^7$ electron filling configuration, which suggests that Co$^{2+}$/Ni$^{3+}$ based materials containing this type of hexagonal lattices are promising new high T$_c$ materials\cite{Hu-tbp}.  However, confirming such a prediction can be very difficult due to the rare appearance of trigonal biprymidal complexes in material databases.

  Here we propose a new family of Co/Ni-based materials that carry the special electronic property to be promising unconventional high-T$_c$ candidates. The materials are constructed by layered square lattices with each layer being formed by vertex-shared tetrahedra cation-anion complexes. When it is close to the $d^7$  filling configurations in the d-shell, namely  those of  Co$^{2+}$ or Ni$^{3+}$,  the electronic physics in these materials are fully attributed to the three near degenerated $t_{2g}$ d-orbitals. The new materials closely resemble both cuprates and iron-based superconductors, and thus can bridge the gap between their electronic properties.  In fact, we predict that the materials have the same d-wave pairing symmetry in superconducting states as cuprates  and  can reach a maximum T$_c$   higher than those of iron-based superconductors.  The new family of materials, if synthesized, can be a fertile new ground to study strongly correlated electronic physics and test various ideas on both cuprates and iron-based superconductors.
 \begin{figure}
\centerline{\includegraphics[width=0.48\textwidth]{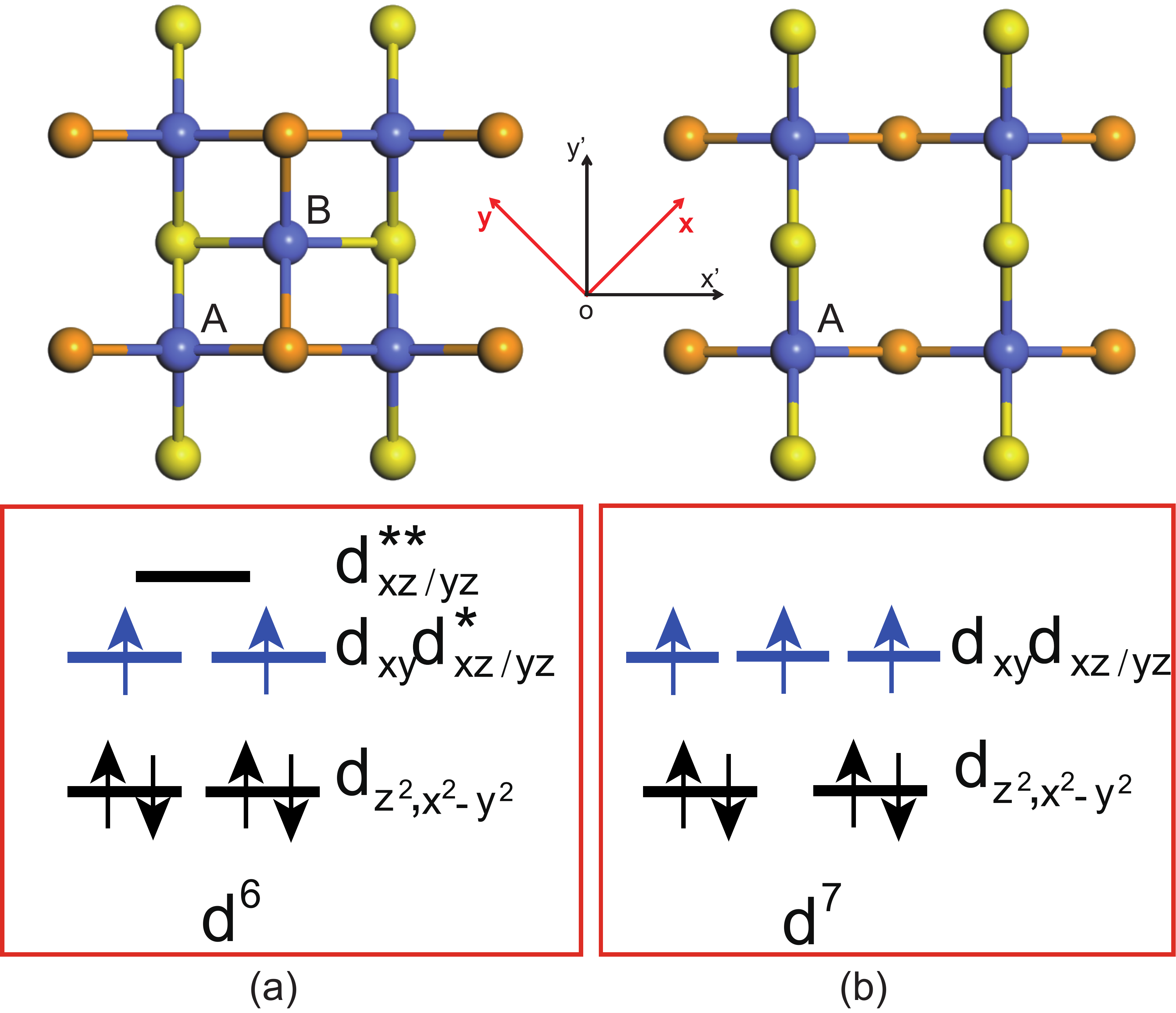}} \caption{(color online) The two dimensional layer structures, the corresponding d-orbital crystal energy splitting configurations and the required electron filling configuration to realize high temperature superconductivity: (a) a FeAs(Se) layer in iron-based superconductors with the d$^6$ filling configuration;(b) the proposed layer by keeping only cation atoms at the A sublattice of (a) with the d$^7$ filling configuration.
 \label{twodimension} }
\end{figure}

 Our proposal is deeply related to iron-based superconductors.  Therefore, we first review the electronic structures of iron-based superconductors. In iron-pnictides, as shown in Fig.\ref{twodimension}(a), the two dimensional FeAs layer is constructed by edge sharing between two nearest neighbour (NN) FeAs$_4$ tetrahedra complexes.   As explicitly pointed out before\cite{Hu-tbp,Hu-genes,Hu2012-s4}, the edge sharing between two sublattices has a profound effect on electronic structures of iron-pnictides. It makes  one combination of $d_{xz}$ and $d_{yz}$ orbitals noted as d$_{xz/yz}^{**}$ strongly couples with the e${_g}$ d$_{x^2-y^2}$ orbital. Such a coupling  creates an energy gap between their binding and antibinding bands to allow the  $d_{xy}$ and the other combination of $d_{xz}$ and $d_{yz}$ orbital noted as d$_{xz/yz}^{*}$,  the two pure t$_{2g}$ orbitals,  separated from all other d-orbitals. As shown in Fig.\ref{twodimension}(a), the $d^6$ filling configuration makes these two pure t$_{2g}$ oribtals isolated near Fermi energy, which explains why $Fe^{2+}$ is specially required to realize high T$_c$ in iron-based superconductors\cite{Hu-genes}. In this analysis, the  superconducting pairing  essentially is confined between two $t_{2g}$ orbitals within each sublattice.

Following the above understanding of iron-based superconductors,  logically we can simply keep one Fe sublattice without losing essential physics. If we divide Fe atoms into two sublattices, each sublattice as shown in Fig.\ref{twodimension}(b)  can be viewed as a structure constructed by vertex sharing between two NN tetrahedra. Thus a natural proposal is to study a material which has a lattice structure of Fig.\ref{twodimension}(b). This is exactly the main point of this paper. It is easy to notice that in this lattice there is no large coupling between e$_g$ and t$_{2g}$ orbitals.  The e$_g$ and t$_{2g}$ orbitals are well separated in energy by crystal field energy splitting. All three t$_{2g}$ orbitals are close to be degenerated. Therefore, as shown in Fig.\ref{twodimension}(b), a configuration close to a $d^7$ filling on transition metal ion meets the requirements to isolate the t$_{2g}$ orbitals near Fermi energy.  Thus, our goal is to construct $Co^{2+}$ or $Ni^{3+}$ based materials containing such a two dimensional lattice structure and predict possible properties.

\begin{figure}
\centerline{\includegraphics[width=0.48\textwidth]{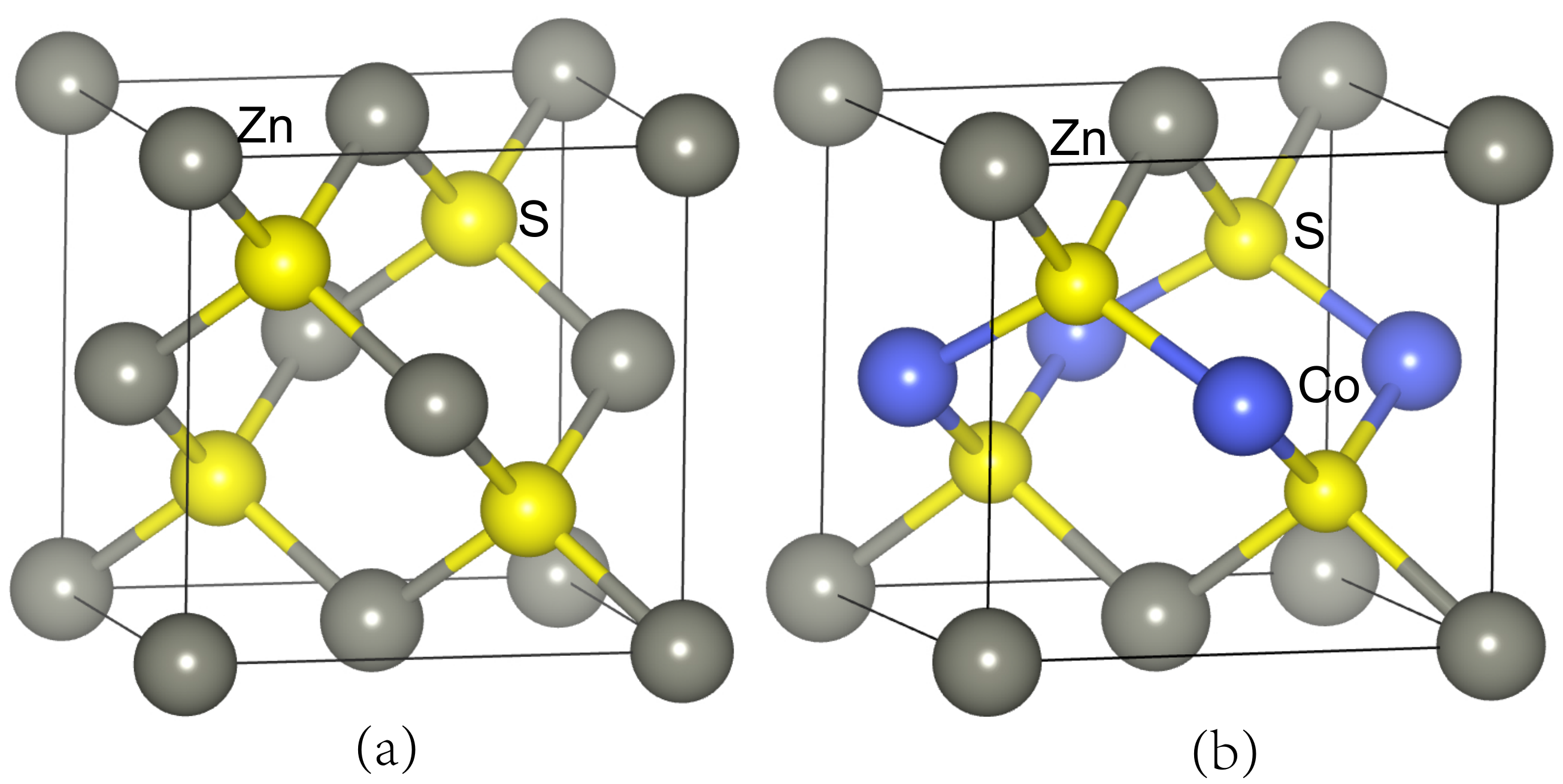}} \caption{(color online) (a) the ZnS(Sphalerite) Zinc Blende structure; (b) a possilbe ZnCoS$_2$ structure with alternating Zn and Co layers.
 \label{zncos2} }
\end{figure}
 The first question is that whether the lattice structure in Fig.\ref{twodimension}(b) is feasible or not. The answer is positive. In fact, the layer structure exists in the popular zinc blende structure($\beta$-ZnS). As shown in Fig.\ref{zncos2}(a), the zinc blende is a well known three dimensional cubic structure created by  vertex-sharing techahedra. If we view the cubic structure layer by layer along any principle axis, each ZnS$_2$ layer is identical to the structure shown in Fig.\ref{twodimension}(b). For the purpose of theoretical demonstration,  we can replace half Zn atoms in $ZnS$ by Co atoms to create a prototype of material ZnCoS$_2$ that has alternating ZnS$_2$ and CoS$_2$ layers along c-axis, shown in Fig.\ref{zncos2}(b).  We can also make extension of this material by replacing S atoms by other chalcogen atoms to create materials such as ZnCoO$_2$ and ZnCoSe$_2$. However, for ZnCoO$_2$, we find that the lattice constant is only about $3.1$\AA, much smaller than those of ZnCo(S,Se)$_2$. The short distance suggests that  there are strong direct hoppings between d-orbitals which can destroy the superexchange processes. Therefore, we will focus on ZnCo(S,Se)$_2$.  In these materials. the electronic physics near Fermi energy is expected to be dominated by the two dimensional Co(S,Se)$_2$  layer as Zn has a filled d-shell.
  \begin{table}[bt]%The best place to locate the table environment is directly after its first reference in text
\caption{\label{structure}%
Optimized structural parameters
of ZnCoS$_2$ and ZnCoSe$_2$ by using GGA .}
%%%% new table
\begin{ruledtabular}
\begin{tabular}{ccc}
%\textrm{Left\footnote{Note a.}}& \textrm{Centered\footnote{Note
%b.}}& \multicolumn{1}{c}{\textrm{Decimal}}&
%\textrm{Right}\\
  & ZnCoS$_2$ & ZnCoSe$_2$ \\
 \colrule
a(\AA) & 3.757 & 3.758  \\
c(\AA) & 5.017 & 6.000  \\
Co-Se(S)(\AA) & 2.172 & 2.319  \\
Se(S)-Co-Se(S) &119.757$^\circ$;104.586$^\circ$&108.204$^\circ$;110.109$^\circ$\\
\end{tabular}
\end{ruledtabular}
\label{table1}
\end{table}

%\textcolor{red}{Our  density functional theory calculations employ the projector augmented
%wave (PAW) method encoded in Vienna ab initio simulation
%package(VASP) \cite{Kresse1993,Kresse1996,Kresse1996B}, and the generalized-gradient approximation
%(GGA) for the exchange correlation functional is used \cite{Perdew1996}.
%The cutoff energy of 500 eV is taken
%for expanding the wave functions into plane-wave basis. In
%the calculation, the number of these k
%points is ($14\times 14 \times7$) for the nonmagnetic calculations and  ($8\times 8 \times8$) for the magnetic calculations.  The GGA plus on-site repulsion U method (GGA+U) in
%the formulation of Dudarev {\it et al}.\cite{Kresse1993,Kresse1996,Kresse1996BDudarev1998} is employed to describe the
%electron correlation effect associated with the Ni/Co $3d$ states by
%an effective parameter $U_{eff}$. The value of $U_{eff}= 4 eV(2 eV)$ on Co is adopted in the calculations. We relax the lattice constants and internal atomic positions with GGA, where the plane wave cutoff energy is 600eV. Forces are minimized to less than 0.01 eV/\AA in the relaxation.}

\begin{figure}
\centerline{\includegraphics[width=0.5\textwidth]{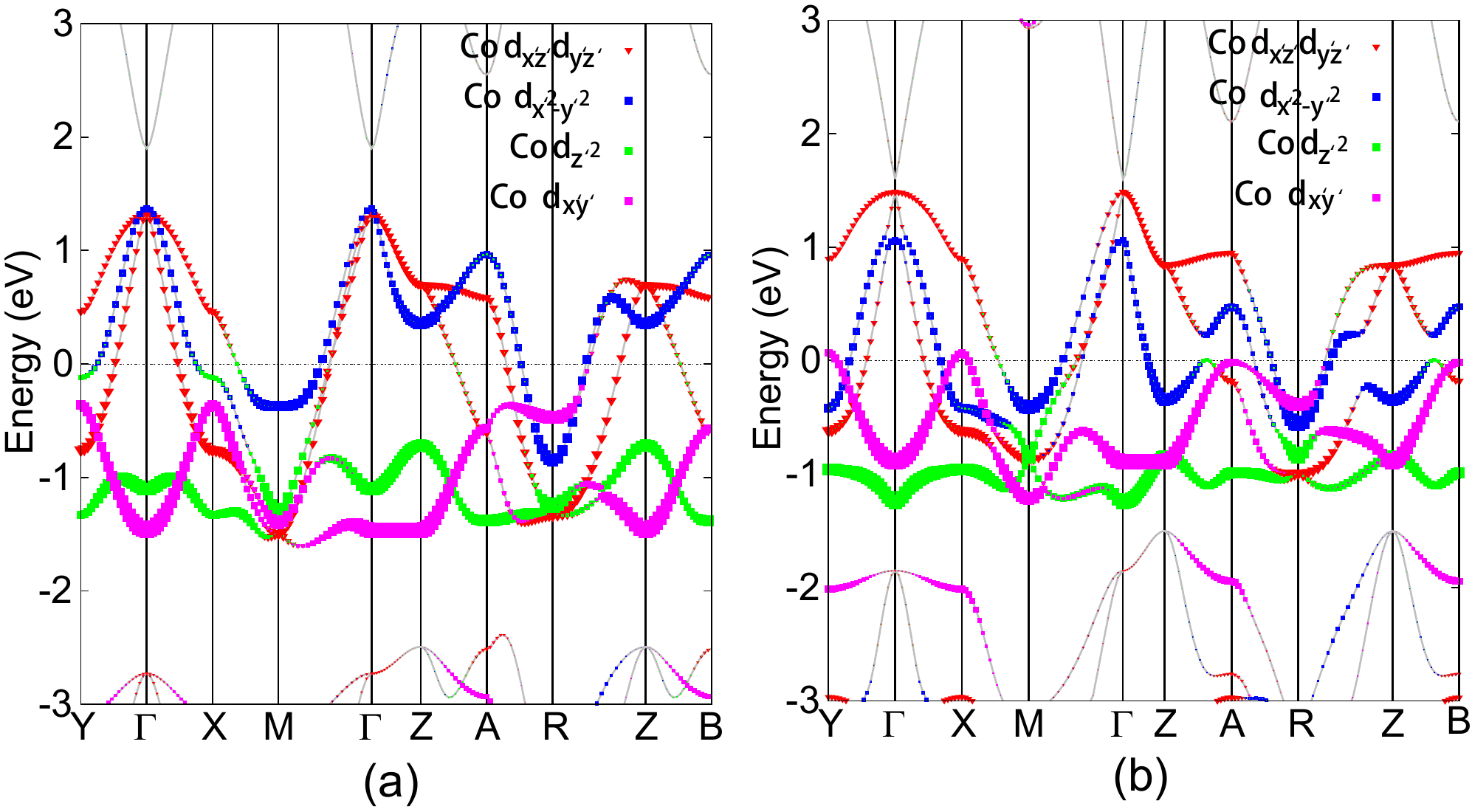}}
\caption{(color online) (a) and (b) are band structures of ZnCoS$_2$ and ZnCoSe$_2$. The orbital characters of bands are represented by different colors.
\label{lattice} }
\label{band}
\end{figure}

We use  the density functional theory (DFT) under the generalized-gradient approximation (GGA) for the exchange correlation functional \cite{Kresse1993,Kresse1996,Kresse1996b,Perdew1996} to obtain the optimized structural parameters for both ZnCo(S,Se)$_2$ which are listed in Table.\ref{table1}. CoSe$_4$ is  close to a perfect tetrahedra. S atoms in ZnCoS$_2$ are much closer to Co layer than Se atoms in ZnCoSe$_2$. These lattice differences are consistent with those between FeS and FeSe in iron-based superconductors\cite{Dagotto2013-review}.
The band structures of these two materials  are shown in Fig.\ref{band} in which different colors mark the orbital characters.  It is very clearly that the three $t_{2g}$ orbitals are  close to half filling and dominate the electronic physics near Fermi energy.  The dispersions along c-axis are not small, which suggests that there are reasonable  couplings between two NN Co layers. This is largely because of  the cubic nature of the original Zinc Blende structure. In principle, we may design a material to make large separations between NN Co layers to reduce the c-axis dispersion. For example, in iron-based superconductors, the c-axis dispersion of the 1111 LaOFeAs is much weaker than those of the 122 BaFe$_2$As$_2$ and the maximum T$_c$ is higher  in the former  than  the latter\cite{Johnston2010-review}.  Therefore, the large c-axis dispersion here is clearly not good for achieving maximum  T$_c$, which suggests  ZnCo(S,Se)$_2$  may be not the optimal materials to  achieve the maximum potential $T_c$ in these families.  Nevertheless, the essential two dimensional physics from the Co(S,Se)$_2$ layer can still be analyzed  in this prototype by focusing on a single two dimensional layer.

\begin{figure}
\centerline{\includegraphics[width=0.5\textwidth]{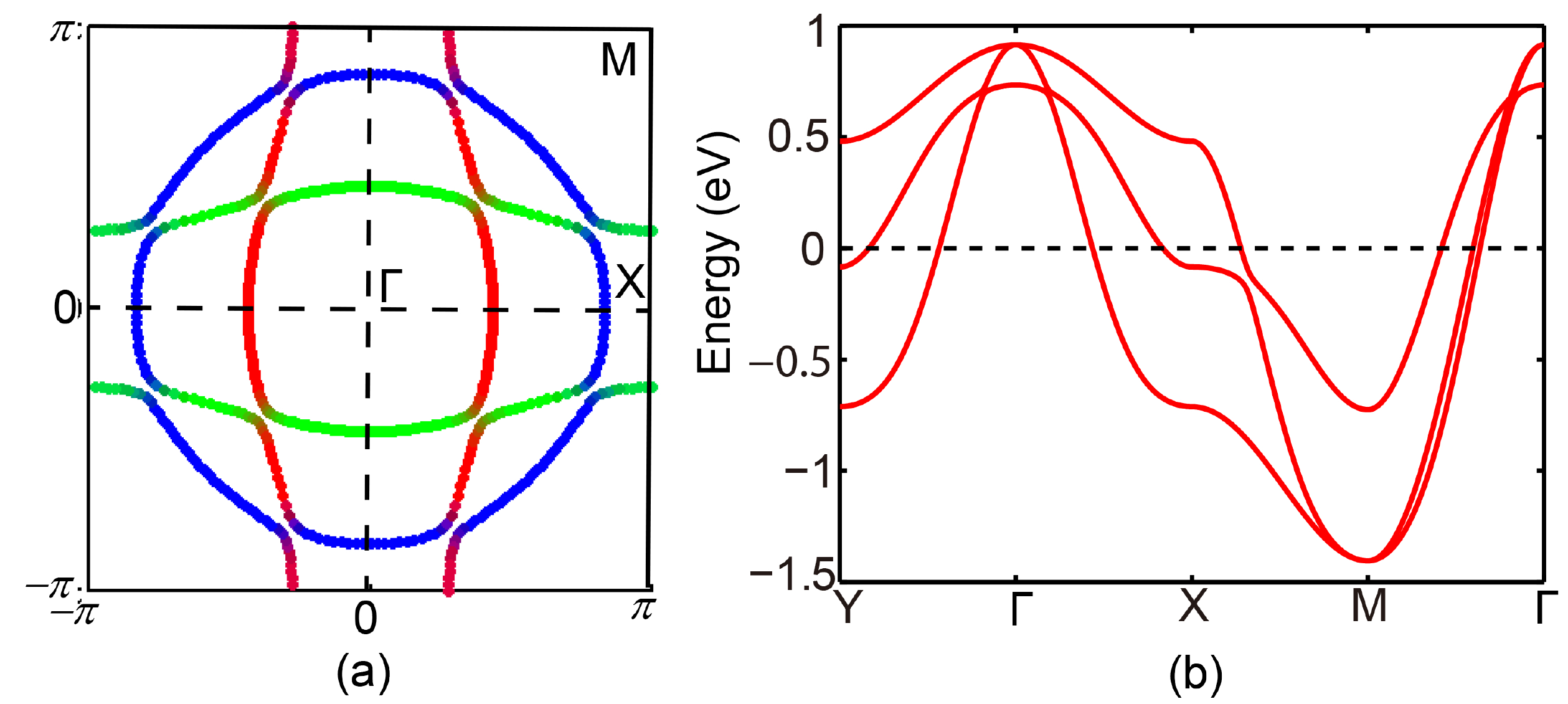}}
\caption{(color online) (a) The fermi surface of ZnCoS$_2$. The orbital contributions
of the different FS sheets are shown color coded: d$_{xz}$(green),d$_{yz}$ (red) and d$_{xy}$ (bule).(b) The band structure of the effective model.
\label{band_fs} }
\end{figure}

The  minimum effective tight binding model Hamiltonian $H_0$ to capture the three t$_{2g}$  orbitals near Fermi surfaces in a single layer  can be written as a $3\times 3$ Hermitian matrix. In the following, we use the  two principle axises of the Co square lattice, $x' $and $y'$ as shown in Fig.\ref{twodimension} and take the t$_{2g}$ orbital base $(d_{x'z}, d_{y'z}, d_{x'^2-y'^2})$. Without causing confusion, we ignore the prime label in the following of the paper.  The elements of $H_0$ matrix are given by
\begin{eqnarray}
\label{eq2}
H_{11} & = & \epsilon_{1}+2t^{11}_{x}cos(k_x) +2t^{11}_{y}cos(k_y)+4t^{11}_{xy}cos(k_x)cos(k_y)\nonumber
\\
&& +2t^{11}_{xx}cos(2k_x) +2t^{11}_{yy}cos(2k_y),\nonumber \\
H_{12} & = & -4t^{12}_{xy}sin(k_x)sin(k_y)\nonumber
\\
H_{13} & = & 2it^{13}_xsin(k_x)+4it^{13}_{xy}sin(k_x)cos(k_y)+2it^{13}_{xx}sin(2k_x)\nonumber
\\
H_{22} & = & \epsilon_{2}+2t^{22}_{x}cos(k_x) +2t^{22}_{y}cos(k_y)+4t^{22}_{xy}cos(k_x)cos(k_y)\nonumber
\\
&& +2t^{22}_{xx}cos(2k_x) +2t^{22}_{yy}cos(2k_y),\nonumber
\\
H_{23} & = & 2it^{23}_ysin(k_y)+4it^{23}_{xy}sin(k_y)cos(k_x)+2it^{23}_{xx}sin(2k_y)
\nonumber
\\
H_{33} & = & \epsilon_{3}+2t^{33}_{x}(cos(k_x)+cos(k_y))+4t^{33}_{xy}cos(k_x)cos(k_y)\nonumber
\\
&& +2t^{33}_{xx}(cos(2k_x)+cos(2k_y))
\end{eqnarray}
We use   eV  as the energy unit for all parameters without further specification.  By fitting to the band structure of ZnCoS$_2$ at the $k_z=0$ plane,
 we have  $\epsilon_{1}=\epsilon_{2}=3.7314$  and $\epsilon_{3}=4.1241$  for the onset energy of $d_{xz,yz}$ and $d_{x^2-y^2}$.   The corresponding hopping parameters in above equation are $t^{11}_{x}=t^{22}_{y}=0.4391$,  $t^{11}_{y}=t^{22}_{x}=0.1408$, $t^{11}_{xy}=t^{11}_{xy}=-0.0162$, $t^{12}_{xy}=0.021$, $t^{13}_x=t^{23}_y=0.0057$,
$t^{13}_{xy}=t^{13}_{xy}=-0.0061$, $t^{33}_x=0.1824$,$t^{33}_{xy}=0.011$, $t^{11}_{xx}=t^{22}_{yy}=0.0688$,
$t^{11}_{yy}=t^{22}_{xx}\textcolor{red}{=}-0.0025$, $t^{13}_{xx}=t^{23}_{yy}=0.0107$, $t^{33}_{xx}=-0.0299$. It is clear that the hoppings are dominated by the NN intra-orbital hoppings. The positive values of these hopping parameters reflect that the hoppings are mediated through the p-orbitals of S/Se anions.  The Fermi surfaces and the  band structure with these parameters are shown in Fig.\ref{band_fs}. The Fermi surfaces are composed of  two large hole pockets around $\Gamma$ and one large electron pockets around $M$.

 Similar to cuprates, here we expect that the electron-electron correlation can be described effectively by onsite interactions.  In this multi-orbital system  it can be written as $H_{eff}= \sum_i H_i$ with
\begin{eqnarray}
H_{i}= \sum_{a} U n_{ia\uparrow}n_{ia\downarrow}+ \sum_{<ab>} (U'n_{ia }n_{ib}- J_H \vec S_{ia}\cdot\vec S_{ib}),
\end{eqnarray}
where the terms represent the Hubbard intra-orbital repulsion, inter-orbital repulsion and Hunds coupling respectively\textcolor{red}{,} and $a,b$ are orbital indices.  Near half filling, as the magnetism is controlled by the AFM superexchanges, the effective Hamiltonian for effective spin-spin interactions  can be written as
\begin{eqnarray}
H_{s}= \sum_{<ij>} J_{ab}  \vec S_{ia}\cdot\vec S_{jb},
\end{eqnarray}
where $<ij>$ is defined between two NN sites. $J_{ab}$ are expected to be dominated by intra-orbital AFM couplings, namely $J_{a=b}$  are much larger than $J_{a\neq b}$.  Combining $H_0$ with these effective interactions provides a minimum model to describe the system.

In this paper, rather than solving above model quantitatively, here we are interested in robust qualitative predictions that can be made for this system based on general principles obtained from cuprates and iron-based superconductors\cite{Hu-genes}. Verifying these predictions can be a strong demonstration of these principles.  In the following, several important issues are in order.

First, we find that the parental compounds of the new system can be a Mott insulator or, at least, is in the vicinity to the Mott insulating phase. The parental compounds of cuprates are Mott insulators but the iron-based superconductors are known to be metallic. These differences have led to many debates whether both materials can be understood  in one category. The new family can serve a bridge  to address this question. On one side,  there are odd number of electrons in the current system, which is the same as the case in cuprates\cite{Anderson2004}. With the odd number, the Mott insulator concept is well defined.  On the other side, the current multi-orbital system  is built on the essential multi-orbital physics of iron-based superconductors.
\begin{figure}
\centerline{\includegraphics[height=5cm]{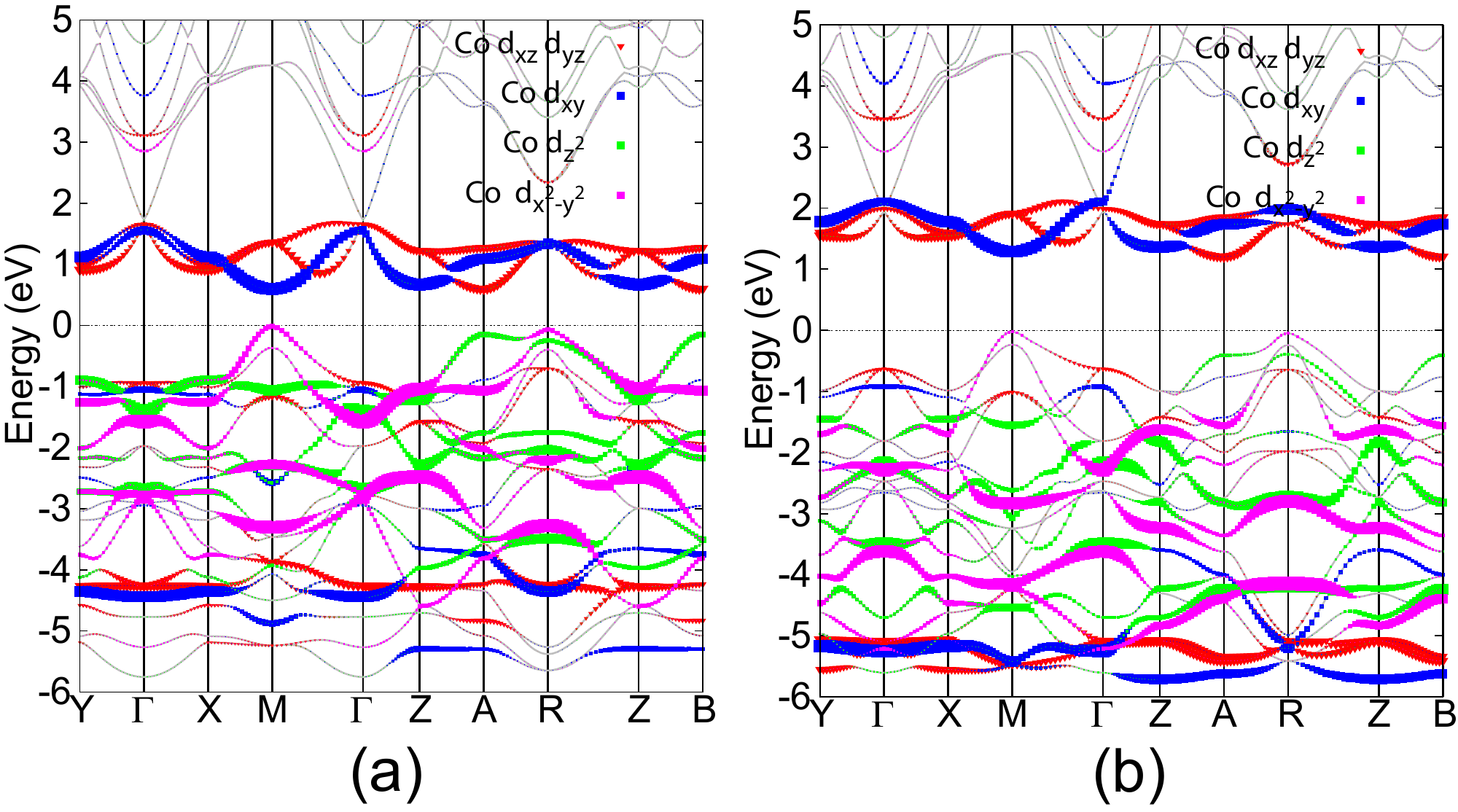}}
\caption{The band structures under the GGA+U calculations for ZnCo$Se_2$: (a) U=2eV; (b) U=4eV.}
 \label{bandU}
\end{figure}
In Fig.\ref{bandU}, we report the GGA+U calculation for ZnCoSe$_2$. The material has a G-type AFM  ground state. The ordered magnetic moment is about $1.76\mu_B$  without U,  $2.14\mu_B$ at  $U=2.0eV$ and $2.37\mu_B$ at $U=4.0eV$. The AFM ground state is  metallic without U.  However, they are  insulating at both $U=2.0eV$ and $U=4.0eV$.  The insulating gap increases as U  increases.  These results are very similar to those from a similar calculation for cuprates and strongly imply for Mottness in this material.

Second, we predict that  in this system, similar to cuprates, the pairing symmetry  in the superconducting state is a robust d-wave pairing upon doping near half filling. Recently, one basic principle to unify the pairing symmetries in cuprates and iron-based superconductors was specified  by Hu and Ding\cite{Huding2012}: the pairing symmetry is simply selected by the overlap between the pairing form factors which  are determined from the short range AFM exchange interactions  and the Fermi surfaces.  A large overlap is also a necessary condition to achieve high T$_c$. This principle, which we refer it as the {\it Hu-Ding} principle, was  also generalized   to include other orders later in ref.\cite{DavisLee2013} by Davis and Lee.  The principle  provides an unified explanation why the d-wave pairing symmetry and the extended s-wave pairing symmetry are robust respectively in curpates and iron-based superconductors\cite{Hu-s-wave}.

In the current system,  as argued earlier, the AFM exchange couplings are dominated by the NN intra-orbital couplings which can generate NN intra-orbital pairing. For a d-wave, the pairing form factor for the d$_{x^2-y^2}$ orbital is $cosk_x-cosk_y$. For the pairing form factors for the d$_{xz}$ and d$_{yz}$ orbitals are  $cosk_x$ and $-cosk_y$ respectively.   These form factors have a very large overlap to the Fermi surfaces shown in Fig.\ref{band_fs}. If we consider the extended s-wave, it is clear that there is little overlap between Fermi surfaces and the extended s-wave form factor $cosk_x+cosk_y$ for d$_{x^2-y^2}$ orbital.  Therefore, following the Hu-Ding principle, the superconducting state should have a robust d-wave pairing symmetry upon doping near half filling.

Finally, we predict that the maximum T$_c$ in these systems should be higher than the maximum T$_c$ achieved in iron-based superconductors.  The superconducting transition temperature can be affected by many factors. However, if the superconducting mechanism is assumed to be identical, we can compare the maximum achievable T$_c$  between different families based on their intrinsic energy scales. The ratio of the  maximum T$_c$s observed in cuprates and iron-based superconductors  is indeed roughly consistent with the energy scale ratio between two families\cite{Hu-tbp}.  In the current system, the energy scale is identical to those of iron-based superconductors. However, because here all t$_{2g}$ orbitals participate in superconducting pairing,  it is reasonable to argue that the maximum T$_c$ should exceed those of iron-based superconductors where only two t$_{2g}$ orbitals essentially make contribution to pairing.

It is interesting to notice that the band structures of our proposed systems are very similar to those of Sr$_2$RuO$_4$ where there are 4 electron in three $t_{2g}$ orbitals\cite{Maeno1994}. However, the physics between these two systems are rathere different. The latter is clearly a weakly correlated electron system and the t$_{2g}$ orbitals are very weakly connected to the surrounding p-orbitals of oxygens.

Our proposed systems can exhibit many unique properties. As the electronic structure is featured with three near degenerated orbitals. It is relatively easy to develop orbital orders. The electronic nematicity which exists in iron-based superconductors can also take place. Moreover,
in this lattice structure, the bonding angle of Co-S(Se)-Co is relatively easy to be changed under pressure. Thus,   many electronic properties can be very sensitive to external or internal pressure.   Combining with Mottness physics,  the interwining physics between all these possible phenomona make this system an extremely intriguing system.

Upon a heavy electron doping, it is also possible that a  superconducting state with a $d\pm is$ pairing symmetry may develop because of the existence of the three t$_{2g}$ orbitals. An extended s-wave for the d$_{xz}$ and d$_{yz}$ is very energetically competitive to the d-wave pairing symmetry based on the Hu-Ding principle, but not for the $d_{x^2-y^2}$ orbital.  Therefore, if the d-wave pairing on the d$_{x^2-y^2}$ orbitals is  weakened upon doping,  it is possible to develop a $d\pm is$ pairing symmetry in which the pairing symmetry within  the d$_{xz}$ and d$_{yz}$ orbitals are an extended s-wave.

In the current material database, we have not found a Co or Ni-based bulk material that carries the proposed two dimensional layer structure. However,   there are Co-based bulk materials with layered Co chains formed by vertex shared  CoSe$_4$, such as CsYbCoSe$_3$\cite{Ibers2007}, in which the valence of the Co atoms are $2+$. The existence of such a stable structure suggests that the proposed structure is feasible. Furthermore, as the   zinc blende structure is a very popular and stable structure, it may be also possible to realize the proposed layer structure as an interface between two zinc blende structure by molecular beam epitaxy(MBE) method\cite{Wang2012-fese}. Finally, although we have focused on S and Se anion atoms in this paper to demonstrate the essential idea, we can also try halogens  or pricogens to check the possibility to form such a desired electronic structure.

In summary, we suggest that a new Co or Ni-based  two dimensional square lattice structure constructed by vertex-shared tetrahedra can realize unconventional high T$_c$ when the electorn filling configuration in the d-shell is close to  d$^7$.   The electronic physics  is fully attributed to the three near degenerated $t_{2g}$ d-orbitals and supports a robust d-wave pairing symmetry. Confirming these predictions can settle the elusive high T$_c$ mechanism regarding both cuprates and iron-based superconductors.

{\it Acknowledgement:}  the work is supported by the Ministry of Science and Technology of China 973 program(Grant No. 2015CB921300), National Science Foundation of China (Grant No. NSFC-11334012), and   the Strategic Priority Research Program of  CAS (Grant No. XDB07000000).

% compare with SR2RUO4
% possible d+is
%lattice distortion and nematicity
%angle dependence, pressure sensitivity
%materials consideration

%\bibliography{genes}

\begin{thebibliography}{18}
\expandafter\ifx\csname natexlab\endcsname\relax\def\natexlab#1{#1}\fi
\expandafter\ifx\csname bibnamefont\endcsname\relax
  \def\bibnamefont#1{#1}\fi
\expandafter\ifx\csname bibfnamefont\endcsname\relax
  \def\bibfnamefont#1{#1}\fi
\expandafter\ifx\csname citenamefont\endcsname\relax
  \def\citenamefont#1{#1}\fi
\expandafter\ifx\csname url\endcsname\relax
  \def\url#1{\texttt{#1}}\fi
\expandafter\ifx\csname urlprefix\endcsname\relax\def\urlprefix{URL }\fi
\providecommand{\bibinfo}[2]{#2}
\providecommand{\eprint}[2][]{\url{#2}}

\bibitem[{\citenamefont{Bednorz and Muller}(1986)}]{Bednorz1986}
\bibinfo{author}{\bibfnamefont{J.~G.} \bibnamefont{Bednorz}} \bibnamefont{and}
  \bibinfo{author}{\bibfnamefont{K.~A.} \bibnamefont{Muller}},
  \bibinfo{journal}{Z. Phys. B} \textbf{\bibinfo{volume}{64}},
  \bibinfo{pages}{189} (\bibinfo{year}{1986}).

\bibitem[{\citenamefont{Kamihara et~al.}(2008)\citenamefont{Kamihara, Watanabe,
  Hirano, and Hosono}}]{Kamihara2008-jacs}
\bibinfo{author}{\bibfnamefont{Y.}~\bibnamefont{Kamihara}},
  \bibinfo{author}{\bibfnamefont{T.}~\bibnamefont{Watanabe}},
  \bibinfo{author}{\bibfnamefont{M.}~\bibnamefont{Hirano}}, \bibnamefont{and}
  \bibinfo{author}{\bibfnamefont{H.}~\bibnamefont{Hosono}},
  \bibinfo{journal}{JACS} \textbf{\bibinfo{volume}{130}}, \bibinfo{pages}{3296}
  (\bibinfo{year}{2008}).

\bibitem[{\citenamefont{Hu et~al.}(2015)\citenamefont{Hu, Le, and Wu}}]{Hu-tbp}
\bibinfo{author}{\bibfnamefont{J.~P.} \bibnamefont{Hu}},
  \bibinfo{author}{\bibfnamefont{C.~C.} \bibnamefont{Le}}, \bibnamefont{and}
  \bibinfo{author}{\bibfnamefont{X.~X.} \bibnamefont{Wu}},
  \bibinfo{journal}{Phys. Rev. X} \textbf{\bibinfo{volume}{5}},
  \bibinfo{pages}{041012} (\bibinfo{year}{2015}).

\bibitem[{\citenamefont{Hu}(2016)}]{Hu-genes}
\bibinfo{author}{\bibfnamefont{J.~P.} \bibnamefont{Hu}},
  \bibinfo{journal}{Science Bulletin} \textbf{\bibinfo{volume}{61}},
  \bibinfo{pages}{561} (\bibinfo{year}{2016}).

\bibitem[{\citenamefont{Hu and Yuan}(2016)}]{Hu-s-wave}
\bibinfo{author}{\bibfnamefont{J.~P.} \bibnamefont{Hu}} \bibnamefont{and}
  \bibinfo{author}{\bibfnamefont{J.}~\bibnamefont{Yuan}},
  \bibinfo{journal}{Front. Phys.} \textbf{\bibinfo{volume}{11}},
  \bibinfo{pages}{117404} (\bibinfo{year}{2016}).

\bibitem[{\citenamefont{Hu and Hao}(2012)}]{Hu2012-s4}
\bibinfo{author}{\bibfnamefont{J.}~\bibnamefont{Hu}} \bibnamefont{and}
  \bibinfo{author}{\bibfnamefont{N.}~\bibnamefont{Hao}},
  \bibinfo{journal}{Phys. Rev X} \textbf{\bibinfo{volume}{2}},
  \bibinfo{pages}{021009} (\bibinfo{year}{2012}).

\bibitem[{\citenamefont{G. and J.}(1993)}]{Kresse1993}
\bibinfo{author}{\bibfnamefont{K.}~\bibnamefont{G.}} \bibnamefont{and}
  \bibinfo{author}{\bibfnamefont{H.}~\bibnamefont{J.}}, \bibinfo{journal}{Phys.
  Rev. B} \textbf{\bibinfo{volume}{47}}, \bibinfo{pages}{558}
  (\bibinfo{year}{1993}).

\bibitem[{\citenamefont{G. and J.}(1996{\natexlab{a}})}]{Kresse1996}
\bibinfo{author}{\bibfnamefont{K.}~\bibnamefont{G.}} \bibnamefont{and}
  \bibinfo{author}{\bibfnamefont{F.}~\bibnamefont{J.}},
  \bibinfo{journal}{Comput. Mater. Sci.} \textbf{\bibinfo{volume}{61}},
  \bibinfo{pages}{15} (\bibinfo{year}{1996}{\natexlab{a}}).

\bibitem[{\citenamefont{G. and J.}(1996{\natexlab{b}})}]{Kresse1996b}
\bibinfo{author}{\bibfnamefont{K.}~\bibnamefont{G.}} \bibnamefont{and}
  \bibinfo{author}{\bibfnamefont{F.}~\bibnamefont{J.}}, \bibinfo{journal}{Phys.
  Rev. B} \textbf{\bibinfo{volume}{54}}, \bibinfo{pages}{11169}
  (\bibinfo{year}{1996}{\natexlab{b}}).

\bibitem[{\citenamefont{Perdew et~al.}(1996)\citenamefont{Perdew, Burke, and
  Ernzerhof}}]{Perdew1996}
\bibinfo{author}{\bibfnamefont{J.}~\bibnamefont{Perdew}},
  \bibinfo{author}{\bibfnamefont{K.}~\bibnamefont{Burke}}, \bibnamefont{and}
  \bibinfo{author}{\bibfnamefont{M.}~\bibnamefont{Ernzerhof}},
  \bibinfo{journal}{Phys. Rev. Lett.} \textbf{\bibinfo{volume}{77}},
  \bibinfo{pages}{3865} (\bibinfo{year}{1996}).

\bibitem[{\citenamefont{Dagotto}(2013)}]{Dagotto2013-review}
\bibinfo{author}{\bibfnamefont{E.}~\bibnamefont{Dagotto}},
  \bibinfo{journal}{Rev. Mod. Phys.} \textbf{\bibinfo{volume}{85}},
  \bibinfo{pages}{849} (\bibinfo{year}{2013}).

\bibitem[{\citenamefont{Johnston}(2010)}]{Johnston2010-review}
\bibinfo{author}{\bibfnamefont{D.~C.} \bibnamefont{Johnston}},
  \bibinfo{journal}{Adv. Phys.} \textbf{\bibinfo{volume}{59}},
  \bibinfo{pages}{803} (\bibinfo{year}{2010}).

\bibitem[{\citenamefont{Anderson et~al.}(2004)\citenamefont{Anderson, Lee,
  Randeria, Rice, Trivedi, and Zhang}}]{Anderson2004}
\bibinfo{author}{\bibfnamefont{P.~W.} \bibnamefont{Anderson}},
  \bibinfo{author}{\bibfnamefont{P.~A.} \bibnamefont{Lee}},
  \bibinfo{author}{\bibfnamefont{M.}~\bibnamefont{Randeria}},
  \bibinfo{author}{\bibfnamefont{T.~M.} \bibnamefont{Rice}},
  \bibinfo{author}{\bibfnamefont{N.}~\bibnamefont{Trivedi}}, \bibnamefont{and}
  \bibinfo{author}{\bibfnamefont{F.~C.} \bibnamefont{Zhang}},
  \bibinfo{journal}{J. Phys.-Cond. Matt.} \textbf{\bibinfo{volume}{16}},
  \bibinfo{pages}{R755} (\bibinfo{year}{2004}).

\bibitem[{\citenamefont{Hu and Ding}(2012)}]{Huding2012}
\bibinfo{author}{\bibfnamefont{J.~P.} \bibnamefont{Hu}} \bibnamefont{and}
  \bibinfo{author}{\bibfnamefont{H.}~\bibnamefont{Ding}},
  \bibinfo{journal}{Scientific Reports} \textbf{\bibinfo{volume}{2}},
  \bibinfo{pages}{381} (\bibinfo{year}{2012}).

\bibitem[{\citenamefont{Davis and Lee}(2013)}]{DavisLee2013}
\bibinfo{author}{\bibfnamefont{J.~S.} \bibnamefont{Davis}} \bibnamefont{and}
  \bibinfo{author}{\bibfnamefont{D.-H.} \bibnamefont{Lee}},
  \bibinfo{journal}{PNAS} \textbf{\bibinfo{volume}{110}},
  \bibinfo{pages}{17623} (\bibinfo{year}{2013}).

\bibitem[{\citenamefont{Maeno et~al.}(1994)\citenamefont{Maeno, Hashimoto,
  Yoshida, Nishizaki, Fujita, Bednorz, and Linchtenberg}}]{Maeno1994}
\bibinfo{author}{\bibfnamefont{Y.}~\bibnamefont{Maeno}},
  \bibinfo{author}{\bibfnamefont{H.}~\bibnamefont{Hashimoto}},
  \bibinfo{author}{\bibfnamefont{K.}~\bibnamefont{Yoshida}},
  \bibinfo{author}{\bibfnamefont{S.}~\bibnamefont{Nishizaki}},
  \bibinfo{author}{\bibfnamefont{T.}~\bibnamefont{Fujita}},
  \bibinfo{author}{\bibfnamefont{J.~G.} \bibnamefont{Bednorz}},
  \bibnamefont{and}
  \bibinfo{author}{\bibfnamefont{F.}~\bibnamefont{Linchtenberg}},
  \bibinfo{journal}{Nature Physics} \textbf{\bibinfo{volume}{372}},
  \bibinfo{pages}{532} (\bibinfo{year}{1994}).

\bibitem[{\citenamefont{Chan et~al.}(2007)\citenamefont{Chan, SHerry, Duyne,
  and Ibers}}]{Ibers2007}
\bibinfo{author}{\bibfnamefont{G.~H.} \bibnamefont{Chan}},
  \bibinfo{author}{\bibfnamefont{L.~J.} \bibnamefont{SHerry}},
  \bibinfo{author}{\bibfnamefont{R.~P.~V.} \bibnamefont{Duyne}},
  \bibnamefont{and} \bibinfo{author}{\bibfnamefont{J.}~\bibnamefont{Ibers}},
  \bibinfo{journal}{Z. Rnorg. Allg. Chem.} \textbf{\bibinfo{volume}{633}},
  \bibinfo{pages}{1343} (\bibinfo{year}{2007}).

\bibitem[{\citenamefont{Wang et~al.}(2012)\citenamefont{Wang, Li, Zhang, Zhang,
  Zhang, Li, Ding, Ou, Deng, Chang et~al.}}]{Wang2012-fese}
\bibinfo{author}{\bibfnamefont{Q.~Y.} \bibnamefont{Wang}},
  \bibinfo{author}{\bibfnamefont{Z.}~\bibnamefont{Li}},
  \bibinfo{author}{\bibfnamefont{W.~H.} \bibnamefont{Zhang}},
  \bibinfo{author}{\bibfnamefont{Z.~C.} \bibnamefont{Zhang}},
  \bibinfo{author}{\bibfnamefont{J.~S.} \bibnamefont{Zhang}},
  \bibinfo{author}{\bibfnamefont{W.}~\bibnamefont{Li}},
  \bibinfo{author}{\bibfnamefont{H.}~\bibnamefont{Ding}},
  \bibinfo{author}{\bibfnamefont{Y.~B.} \bibnamefont{Ou}},
  \bibinfo{author}{\bibfnamefont{P.}~\bibnamefont{Deng}},
  \bibinfo{author}{\bibfnamefont{K.}~\bibnamefont{Chang}},
  \bibnamefont{et~al.}, \bibinfo{journal}{Chin. Phys. Lett.}
  \textbf{\bibinfo{volume}{29}}, \bibinfo{pages}{037402}
  (\bibinfo{year}{2012}).

\end{thebibliography}

%\begin{thebibliography}{K2Cr3As3_pairing}
%%\bibitem{Hu2012} J. P. Hu and H. Ding, Sci. Rep. {\bf 2}, 381 (2012).
%%%DFT
%\bibitem{Kresse1993} G. Kresse and J. Hafner, Phys. Rev. B {\bf 47}, 558 (1993).
%\bibitem{Kresse1996} G. Kresse and J. Furthmuller, Comput. Mater. Sci. {\bf 6}, 15 (1996).
%\bibitem{Kresse1996B} G. Kresse and J. Furthmuller, Phys. Rev. B {\bf 54}, 11169 (1996).
%\bibitem{Perdew1996} J. P. Perdew, K. Burke, and M. Ernzerhof, Phys. Rev. Lett. {\bf 77},
%3865 (1996).
%
%\end{thebibliography}

%\begin{thebibliography}
%\bibitem
%\end{thebibliography}

\end{document}